\documentclass[twocolumn,prl,aps,superscriptaddress,nobalancelastpage]{revtex4}
\usepackage{graphicx,bm}
\usepackage{times}
\begin{document}

\title{Quasiparticle mass enhancement close to the quantum critical point in BaFe$_2$(As$_{1-x}$P$_x$)$_2$}

\author{P. Walmsley}
\affiliation{H. H. Wills Physics Laboratory, University of Bristol, Tyndall Avenue, Bristol, BS8 1TL, United Kingdom.}
\author{C. Putzke}
\affiliation{H. H. Wills Physics Laboratory, University of Bristol, Tyndall Avenue, Bristol, BS8 1TL, United Kingdom.}
\author{L. Malone}
\affiliation{H. H. Wills Physics Laboratory, University of Bristol, Tyndall Avenue, Bristol, BS8 1TL, United Kingdom.}
\author{I. Guillam\'{o}n}
\affiliation{H. H. Wills Physics Laboratory, University of Bristol, Tyndall Avenue, Bristol, BS8 1TL, United Kingdom.}
\author{D. Vignolles}
\affiliation{Laboratoire National des Champs Magn\'{e}tiques Intenses (CNRS-INSA-UJF-UPS), 31400 Toulouse, France.}
\author{C. Proust}
\affiliation{Laboratoire National des Champs Magn\'{e}tiques Intenses (CNRS-INSA-UJF-UPS), 31400 Toulouse, France.}
\author{S. Badoux}
\affiliation{Laboratoire National des Champs Magn\'{e}tiques Intenses (CNRS-INSA-UJF-UPS), 31400 Toulouse, France.}
\author{A.I. Coldea}
\affiliation{Clarendon Laboratory, Department of Physics, University of Oxford, Oxford OX1 3PU, United Kingdom.}
\author{M.D. Watson}
\affiliation{Clarendon Laboratory, Department of Physics, University of Oxford, Oxford OX1 3PU, United Kingdom.}
\author{S. Kasahara}
\affiliation{Department of Physics, Kyoto University, Sakyo-ku, Kyoto 606-8502, Japan.}
\author{Y. Mizukami}
\affiliation{Department of Physics, Kyoto University, Sakyo-ku, Kyoto 606-8502, Japan.}
\author{T. Shibauchi}
\affiliation{Department of Physics, Kyoto University, Sakyo-ku, Kyoto 606-8502, Japan.}
\author{Y. Matsuda}
\affiliation{Department of Physics, Kyoto University, Sakyo-ku, Kyoto 606-8502, Japan.}
\author{A. Carrington}
\affiliation{H. H. Wills Physics Laboratory, University of Bristol, Tyndall Avenue, Bristol, BS8 1TL, United Kingdom.}

\begin{abstract}
We report a combined study of the specific heat and de Haas-van Alphen effect in the iron-pnictide superconductor BaFe$_2$(As$_{1-x}$P$_x$)$_2$.  Our data when combined with results for the magnetic penetration depth give compelling evidence for the existence of a quantum critical point (QCP) close to $x=0.30$ which affects the majority of the Fermi surface by enhancing the quasiparticle mass.  The results show that the sharp peak in the inverse superfluid density seen in this system results from a strong increase in the quasiparticle mass at the QCP.
\end{abstract}

\maketitle
Recently there has been much interest in the idea that superconductivity in the iron-pnictides is boosted by the presence of a quantum critical point (QCP) located at the zero temperature end point of an antiferromagnetic phase transition \cite{Abrahams11}.  It has been suggested that the presence of fluctuations close to this point increases the superconducting transition temperature by increasing the strength of the pairing interactions and / or by increasing the energy of the normal state relative to the superconducting state.

In most iron-pnictides experimental evidence for a QCP has been difficult to find as its effects are easily masked by inhomogeneity and impurity scattering. BaFe$_2$(As$_{1-x}$P$_x$)$_2$ provides a particularly clean system in which to study these effects as the substitution of As by the isovalent ion P suppresses antiferromagnetism and induces superconductivity \cite{Jian09} without changing the electron/hole balance and without inducing appreciable scattering \cite{Shishido10,Beek10}. These are likely the main reasons why the effect of the QCP on the physical properties are so clearly visible experimentally in  BaFe$_2$(As$_{1-x}$P$_x$)$_2$ but not in its charge doped counterparts.  Experimental evidence for a QCP in this system at $x_c \simeq 0.3$  include a funnel of $T$-linear behavior in the resistivity centered on $x_c$ \cite{Kasahara10} and an increase in effective mass on one of the electron Fermi surfaces, as measured by the de Haas-van Alphen (dHvA) effect \cite{Shishido10}, as $x$ approaches $x_c$.  Nuclear magnetic resonance (NMR) experiments show that the magnetic ordering temperature approaches zero at $x_c$ \cite{Nakai10}.  Most recently, measurements of the magnetic penetration depth $\lambda$ showed a sharp peak in $\lambda$ at $x_c$ indicating that the superfluid density is minimal at this critical $x$ value and therefore that the QCP survives under the superconducting dome \cite{Hashimoto12}. This behavior of $\lambda$ in  BaFe$_2$(As$_{1-x}$P$_x$)$_2$ is so far unique amongst the various putative quantum critical superconducting systems including other iron-pnictides, heavy fermions and cuprates \cite{Shibauchi2013}.

In spite of the above, there is little thermodynamic evidence for the QCP and it is unclear if the peak in $\lambda$ at $x_c$ originates from quasiparticle mass renormalization \cite{Levchenko1212.5719} or other factors such as a reduction of the Fermi volume, phase-fluctuations or even mesoscopic inhomogeneity.  Furthermore if mass renormalization is the main factor, there is a question as to how this varies with temperature or magnetic field or electron momentum (Fermi surface sheet).  Here we report a combined study of the specific heat and dHvA effect of BaFe$_2$(As$_{1-x}$P$_x$)$_2$ which addresses these questions.

\begin{figure}
\center
\includegraphics[width=8cm]{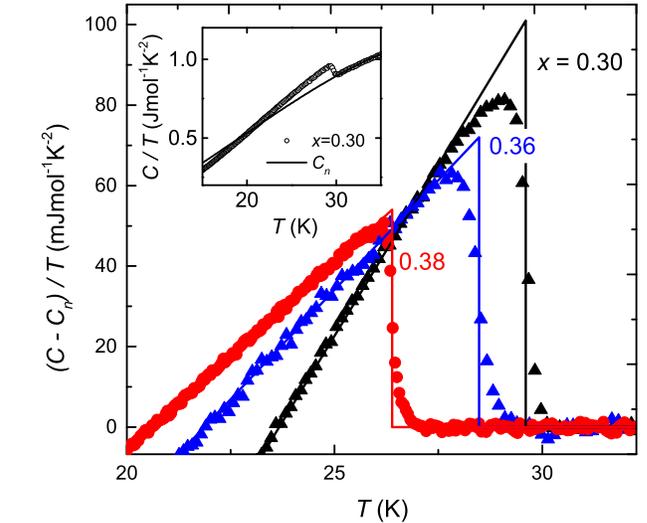}
\caption{(color online) The inset shows the total measured specific heat for a sample of BaFe$_2$(As$_{1-x}$P$_x$)$_2$ with $x=0.30$, the solid line is the fitted normal state background $C_n$.  The main part of the figure shows the specific heat with the normal state $C_n$ subtracted for different values of $x$ and the solid lines show the entropy conserving construction used to determine the jump height $\Delta C$ and $T_c$.} \label{Fig:cpraw}
\end{figure}

\begin{figure}
\center
\includegraphics[width=8cm]{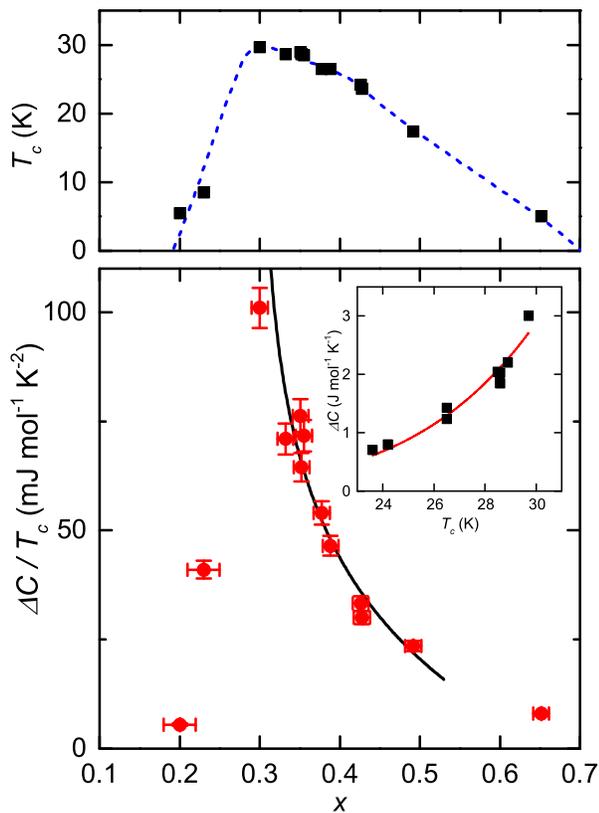}
\caption{(color online) The size of the jump in the specific heat $\Delta C/T_c$ as a function of $x$.  The line shows a fit to the logarithmic behavior expected near a QCP. The inset shows $\Delta C$ vs $T_c$, with the solid line showing a fit to $\Delta C \propto T^n$. The $x$ dependence of $T_c$ is shown in the uppermost panel for the same samples and the dashed line is a guide to the eye.} \label{Fig:cpjump}
\end{figure}

Single crystals of BaFe$_2$(As$_{1-x}$P$_x$)$_2$ were grown by a self-flux technique as described previously \cite{Kasahara10}. A particular requirement of both our specific heat and dHvA studies was to have very homogeneous samples.  To this end we mostly focused on using small crystals with masses 4--70\,$\mu$g for specific heat and 0.1--4\,$\mu$g for dHvA.  To measure the specific heat of such small samples we used a custom build microcalorimeter, where the thermometry elements and heater are deposited as thin films on a SiN membrane.  Calibration of this device was checked by measuring high purity samples of Ag.  We also used a more conventional long relaxation calorimeter \cite{Taylor07} for some larger samples (of order 300\,$\mu$g).   In total around 30 samples were measured using the microcalorimeter and only those with the very sharpest superconducting transitions were retained for further analysis.  For these samples, we made careful x-ray measurements to determine accurately the lattice constants and then used Vegard's law in conjunction with the lattice parameters for BaFe$_2$As$_2$ and BaFe$_2$P$_2$ to determine the value of $x$. dHvA measurements were performed by the torque method using micro piezoresistive cantilevers in both pulsed field (up to 60\,T in Toulouse) and static field (up to 45\,T in Tallahassee). Density functional theory (DFT) band-structure calculations were performed using Wien2K as described previously \cite{Arnold11}.

In Fig.\ \ref{Fig:cpraw} we show the jump in the specific heat $C$ at $T_c$ for samples with $x$ close to $x_c$. Here the anomaly at the transition $\Delta C$ has been isolated from the phonon dominated background by subtracting a  second order polynomial fitted above $T_c$ and extrapolated to lower temperature. Although there would be some uncertainty in using this procedure over an extended temperature range, the lack of appreciable thermal superconducting fluctuations, as evidenced by the mean-field-like form of the anomaly, means that there is very little uncertainty in the size of $\Delta C$.

It is evident from the data in Fig.\ \ref{Fig:cpraw} that the size of the anomaly $\Delta C/T_c$ depends very strongly on $x$ and $T_c$. This is shown in more detail in Fig.\ \ref{Fig:cpjump}. As the strong increase in $\Delta C$ with $x$  is accompanied by a relatively small increase in $T_c$ close to the critical point it seems highly unlikely that the coupling strength is the main factor.   Indeed the normalized slope of $C(T)-C_n(T)$ just below $T_c$, $(T_c/\Delta C)d(C-C_n)/dT$, which depends strongly on the coupling strength \cite{Carbotte90}, is almost independent of $x$ for $0.3<x<0.5$ \cite{slopenote}.  Instead we suggest that this increase in $\Delta C$ reflects the increase in the normal state density of states and hence the Sommerfeld coefficient $\gamma$. We show below that this interpretation gives remarkably good agreement with measurements of the quasiparticle mass from other techniques. We find that our data, for $0.5\gtrsim x \gtrsim 0.3$, are well described by the logarithmic critical behavior, $\Delta C/T_c = c_0+c_1\ln(x-x_c)$, expected close to a QCP \cite{Abrahams11}.

\begin{figure}
\center
\includegraphics[width=8cm]{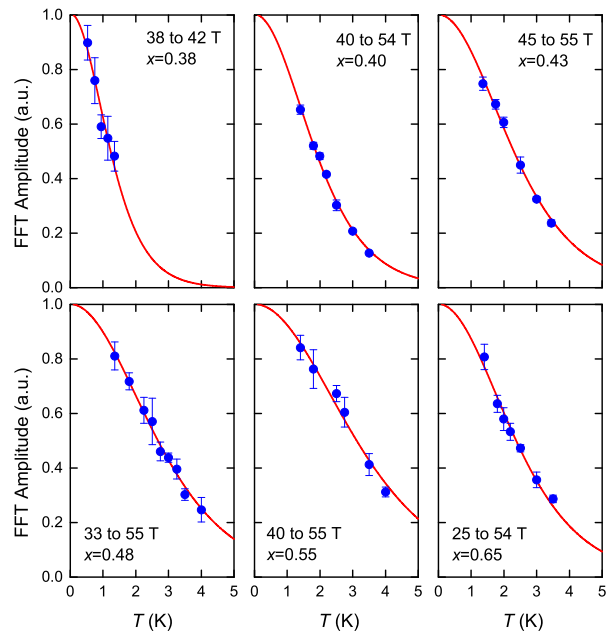}
\caption{(color online) Temperature dependent amplitude of fast Fourier transform (FFT) of the dHvA signal for the $\beta$ orbits for samples with different value of $x$.  The lines are fits to the LK formula. The field range and temperatures are indicated in the plots.} \label{Fig:dhvaraw}
\end{figure}

A strong increase in $\Delta C$ with $T_c$ has been observed previously in many different iron based superconductors including BaFe$_2$(As$_{1-x}$P$_x$)$_2$ \cite{Budko09,Chaparro12}.   Bud'ko, Ni and Canfield (BNC) \cite{Budko09} found that the data for materials with a wide range of $T_c$ could be described by the scaling law $\Delta C \propto T_c^3$.  This has been interpreted as either originating from quantum critically \cite{Zaanen09} or from strong impurity pair breaking\cite{Kogan10}.  In our samples we find significant departures from this scaling and close to the critical doping a much stronger dependence on $T_c$ is observed; a fit to the power law $\Delta C\propto T_c^n$ gives $n=6.5\pm0.7$ for 30\,K$>T_c>$23\,K (inset Fig.\ \ref{Fig:cpjump}).

\begin{figure}
\center
\includegraphics[width=8cm]{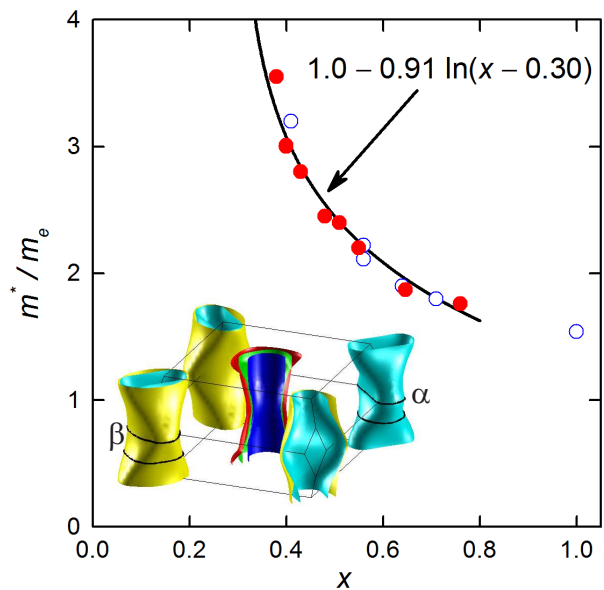}
\caption{(color online) Dependence of the measured dHvA mass of the $\beta$ orbits on $x$, the solid symbols are the new data from the present study and the open symbols are from our previous studies \cite{Shishido10,Arnold11}.  The solid line is a fit to the logarithmic behavior indicated (the point at $x=1$ was not included in the fit).  The inset shows the calculated DFT Fermi surface for BaFe$_2$As$_2$ with lattice parameters and internal positions appropriate to  BaFe$_2$(As$_{1-x}$P$_x$)$_2$ with $x=0.4$. The $\alpha$ and $\beta$ extremal orbits are indicated.} \label{Fig:dhvamassx}
\end{figure}

Although our absolute values of $C$ (including phonons) are similar to those reported in Ref.\ \cite{Chaparro12}, for samples with the highest $T_c$ we find values of $\Delta C$ which are up to a factor of two larger.  The values of $\Delta C$ are very similar for the low $T_c$ samples. This difference likely arises from the higher homogeneity of the present samples.  Indeed, we observed a similar reduction in $\Delta C$ for samples with broader transitions. The sharply peaked behavior of $\Delta C$ and strong asymmetry of $T_c(x)$ close to $x_c$ (Fig.\ \ref{Fig:cpjump}) naturally leads to a strong suppression of $\Delta C$ as the distribution of $x$ within a sample becomes broader.

Specific heat measures the total density of the states (DOS) and so contains contributions from all the Fermi surface sheets. According to density functional theory calculations \cite{Arnold11} for the end member BaFe$_2$P$_2$ the contributions from each sheet to the total DOS is 23\% (band 1 hole), 37\% (band 2 hole), 20\% (band 3 electron), 20\% (band 4 electron).  So the contribution from the holes is a little higher than from the electrons but they are roughly equal.  dHvA effect measurements have the potential to resolve quasiparticle masses from individual orbits on each Fermi surface sheet.  For BaFe$_2$P$_2$ masses of almost all the observable orbits were reported by Arnold \emph{et al.} \cite{Arnold11}. All orbits had relatively uniform mass enhancements $m^*/m_b$ ranging from 1.6 to 1.9.  For the As substituted samples, the signals from the hole sheets are rapidly attenuated as $x$ is decreased, but the signals from the inner ($\alpha$) and outer ($\beta$) electron sheets less so \cite{Shishido10}. The $\beta$ orbits are the most prominent and can be tracked to the lowest value of $x$ and hence highest $T_c$.   In Fig.\ \ref{Fig:dhvaraw} we show how the amplitude of the dHvA signal from the $\beta$ orbit reduces with temperature for several different values of $x$, along with fits to the standard Lifshitz-Kosevich (LK) formula \cite{Shoenberg} which is used to determine the orbitally averaged effective mass $m^*_\beta$.

In Fig.\ \ref{Fig:dhvamassx} we show the effective mass of these orbits over a wide range of doping, up to $x=0.38$, $T_c=28$\,K which is close to the maximum $T_c$.  For the lower values of $x$ we are not able to resolve the maximal and minimal $\beta$ orbits separately because of the restricted range of inverse field over which the signal is observable, so this mass represents an average of the two.  In this figure $x$ was determined using the measured dHvA $\beta$ frequency as this varies linearly with $x$ \cite{Shishido10} and this is more precise than x-ray diffraction.  The solid line in Fig.\ \ref{Fig:dhvamassx} is a fit to the logarithmic form used for the specific heat.
\begin{figure}
\center
\includegraphics*[width=8cm]{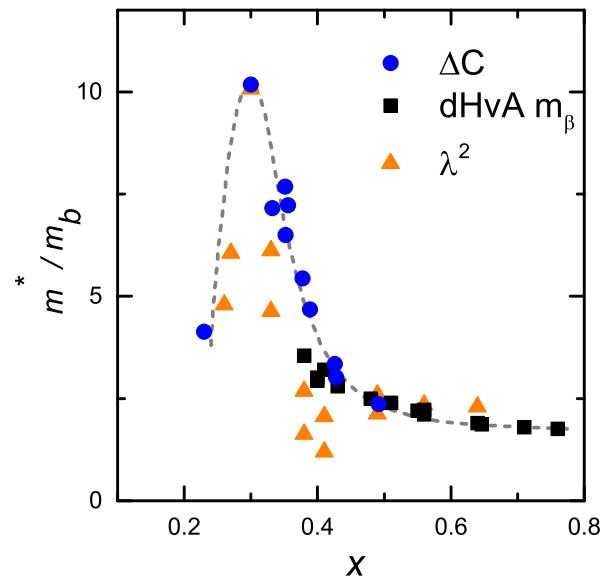}
\caption{(color online) Mass enhancements as derived from specific heat, dHvA and microwave magnetic penetration depth measurements\cite{Hashimoto12}.  The dashed line is a guide to the eye.} \label{Fig:combinedmass}
\end{figure}

We now make a quantitative comparison between the different measurements of effective mass.  For the specific heat data, we first calculate $\gamma$ from the measured values of $\Delta C/T_c$, using the relation $\alpha_c=\Delta C/\gamma T_c$. Although $\alpha_c$ does vary with the strength of the electron-boson coupling and the anisotropy of the superconducting gap, it only increases by a factor of $\sim$2 even for extremely strongly coupled superconductors like Pb \cite{Carbotte90}.  Gap anisotropy decreases $\alpha_c$, for example for $d$-wave $\alpha_c$ is 0.66 of the isotropic $s$-wave value.  As both these effects are relatively small compared to the changes in $\Delta C(x)$ and will cancel each other out to some extent in BaFe$_2$(As$_{1-x}$P$_x$)$_2$ which has gap nodes \cite{Hashimoto10}, we make the approximation that $\alpha_c$ takes the weak coupling $s$-wave value of $\alpha_c=1.43$ for all $x$. Then we calculate an average effective mass enhancement by taking the ratio of $\gamma$ to the band-structure value calculated by DFT for BaFe$_2$P$_2$, $\gamma_b=6.94$\,mJ mol$^{-1}$K$^{-2}$.   $\Delta C$ will be reduced at the edge of superconducting dome because of pair-breaking and sample inhomogeneity so we would not expect the samples with $x=0.2$ and $x=0.65$ to provide a good estimate of $m^*$.

For the penetration depth data we make a similar comparison to the DFT calculations. Using the experimental values of $\lambda_0$  at low temperature measured by the microwave cavity method \cite{Hashimoto12}, we estimate the renormalization from $m^*_\lambda/m_b=A\lambda^2_0/\lambda^2_b$. Here $A=n_{dHvA}/n_b=0.358+0.44x$ is the measured shrinkage of the Fermi surface volume with $x$ \cite{Shishido10} divided by the DFT volume for $x=1$ and $\lambda_b=660$\,\AA~ is the DFT($x=1$) value of $\lambda$. Although $A$ is estimated from the electron sheet ($\alpha$ and $\beta$) dHvA data only, the hole sheets must shrink similarly because the material remains compensated for all $x$.

Finally, for the dHvA data we simply plot the measured values of $m^*_\beta/m_b$ for the $\beta$ orbits. We use the ($x=1$) DFT $m_b$ and $\gamma_b$ values for all $x$ for simplicity as these do not change appreciably with band energy shift corresponding to our range of $x$.  The band-structure of BaFe$_2$(As$_{1-x}$P$_x$)$_2$ is close to the 2D limit where $\gamma$ is directly proportional to $m^*$ and independent of $n$.

In Fig.\ \ref{Fig:combinedmass} it can be seen that these different estimates of the effective mass are remarkably consistent over most of the phase diagram.  This might not be expected for a number of reasons.  First, as $\Delta C/T_c$ is proportional to the thermodynamic mass at $T_c$ and $\lambda^2$ is proportional to the dynamic mass at $T=0$ our result suggests that these masses are very similar. This implies that for $T\lesssim T_c$ we are in an enhanced Fermi liquid regime for all values of $x$ where $\gamma$ is enhanced by the fluctuations but independent of $T$ \cite{Oeschler08}. This is surprising because the resistivity shows a non-Fermi liquid $T$-linear behavior for $T>T_c$ at the same values of $x$. The large $B$ field used for the dHvA measurements also does not seem to decrease the mass except perhaps in the highest $T_c$ sample measurable by dHvA ($x=0.38$).  Second, in a one-component Galilean invariant Fermi liquid, enhancement of $\lambda^2$ is not expected because of `backflow' cancellation \cite{Leggett65}. The agreement we see suggests that this cancellation does not occur in multiband iron-pnictides as suggested by recent theory \cite{Levchenko1212.5719}.  Third, it shows that the mass is uniformly enhanced on all the Fermi surface sheets. In the quasi-classical theory the superfluid density  $\lambda_x^{-2}\propto \int  v_x^2|\bm{v}|^{-1}d\bm{S}$, and so the light electrons where the Fermi velocity $\bm{v}$ is high make the largest contribution. On the other hand, for the specific heat $\gamma\propto \int |\bm{v}|^{-1}d\bm{S} $ and so the heavy electrons contribute most.  Hence the quantitative agreement between the trends for the mass enhancements indicated by $\lambda^2$, $\Delta C$  and the dHvA results for the electron $\beta$ orbits tend to suggest that the mass enhancement is quite uniform over the Fermi surface. This does not exclude the presence of `hot spots' or localized regions with much higher enhancement, which might result from regions that are particularly well connected by strong spin-fluctuation modes, but does suggest rather similar average enhancement for the electron and hole Fermi surfaces. This would be expected if the enhancement results from spin-fluctuated mediated scattering from the electron to hole surfaces with similar orbital character.

In summary, combined data for specific heat, de Haas-van Alphen effect and magnetic penetration depth \cite{Hashimoto12} shows compelling evidence for the existence of a quantum critical point close to $x=0.30$ in the BaFe$_2$(As$_{1-x}$P$_x$)$_2$ system, and that this effects the majority of the Fermi surface by enhancing the quasiparticle mass.  The results show that the sharp peak in the inverse superfluid density seen in this system results from a strong enhancement of the quasiparticle mass at the QCP. The enhanced quasiparticle mass implies that the Fermi energy is suppressed, which is usually less advantageous for high $T_c$. The fact that the highest $T_c$ is nevertheless attained right at $x_c=0.3$ with the most enhanced mass strongly suggests that the quantum critical fluctuations help to enhance superconductivity in this system.

This work was supported by the EPSRC (UK), EuroMagNET II under the EU Contract No. 228043, and KAKENHI from JSPS.  A portion of this work was performed at the National High Magnetic Field Laboratory, which is supported by National Science Foundation Cooperative Agreement No. DMR-0654118, the State of Florida, and the U.S. Department of Energy. I.G. acknowledges Marie Curie Intra-European Fellowship support under contract no. FP7-PEOPLE-2010-IEF-273105.

\bibliographystyle{apsrevNOETAL}
\bibliography{Ba122CdHvaQC}

\begin{thebibliography}{21}
\bibitem[{slo()}]{slopenote}
\bibinfo{note}{Although $(T_c/\Delta C)d(C-C_n)/dT$ is quite sensitive to the
  background subtraction, particularly for samples with small $\Delta C$
  anomalies.}
\end{thebibliography}

\end{document}